\documentclass[aps,prd,tightenlines,preprint,nofootinbib,a4paper]{revtex4}
\usepackage{epsfig}

\begin{document}

\title{Renormalization of the Neutrino Mass Matrix}

\author{S. H. Chiu\footnote{schiu@mail.cgu.edu.tw}}
\affiliation{Physics Group, CGE, Chang Gung University, 
Taoyuan 33302, Taiwan}

\author{T. K. Kuo\footnote{tkkuo@purdue.edu}}
\affiliation{Department of Physics, Purdue University, West Lafayette, IN 47907, USA}

\begin{abstract}

In terms of a rephasing invariant parametrization, the set of renormalization group equations
(RGE) for Dirac neutrino parameters can be cast in a compact and simple form.
These equations exhibit manifest symmetry under flavor permutations.
We obtain both exact and approximate RGE invariants, in addition to some
approximate solutions and examples of numerical solutions.

\end{abstract}


\maketitle

\pagenumbering{arabic}



\section{Introduction}

Despite the tremendous progress made recently regarding the intrinsic properties of
the neutrinos, much remains to be learned.  For instance, it is not known whether the neutrinos
are Dirac or Majorana particles.  The pattern of masses and mixing parameters is also a mystery.
Within the framework of the standard model (SM), it is well-established that
these parameters are functions of the energy scale, and are governed by the 
renormalization group equations (RGE).  Indeed, the evolution of the gauge couplings 
(e.g., asymptotic freedom and gauge coupling unification) 
has been very successfully applied to particle physics, forming an essential part of the 
foundation of the standard model.  One would expect similar considerations to apply
to the Higgs couplings (i.e., the mass matrices).
In addition, these effects are indispensable toward a theoretical understanding of the mass matrices.
Unfortunately, to this date existing results are usually numerical in nature and not very general.
The main difficulty lies in the complexity of the RGE for these parameters.
While the RGE in terms of the mass matrices are reasonably simple, 
these matrices contain a large number of unphysical degrees of freedom, 
and the RGE for physical variables are very complicated indeed.  For instance,
the mixing matrix is rephasing invariant, which is characteristic of the mixing
of quantum states, for which the phase of an individual state vector is unobservable.
To extract the physical parameters, one may proceed by fixing some phases.
This is similar to ``fixing a gauge" in gauge theories.  Thus, in the standard 
parametrization, one sets four phases in the mixing matrix (those of
$V_{11}$, $V_{12}$, $V_{23}$, $V_{33}$) to vanish, arriving at three angles and a phase.
When applied to the analysis of the RGE, one is enforcing this ``gauge" at every
energy scale.  The lack of a theoretical rationale for this choice, we believe, 
contributes to the complexity of the RGE, when expressed in terms of the standard parameters.

In this work we investigate the one-loop RGE evolution of Dirac neutrinos, using a 
rephasing invariant parametrization introduced earlier.
What characterizes this parametrization is its symmetry structure under flavor permutations.
Indeed, we establish a set of RGE which exhibits a simple structure with built-in
permutation symmetry.  This set of equations is not as formidable as the one written
in terms of the standard parametrization. Its solutions are studied both analytically
and numerically.  We found a RGE invariant, in addition to
obtaining some approximate solutions. Numerical examples are also presented.

This work is organized as follows.
In Section II, we briefly introduce the rephasing invariant parametrization 
that will be used in this work. Their properties are also discussed.
In Section III, 
we obtain the RGE in a compact and simple form.  From these we derive both 
exact and approximate RGE invariants in Section IV. 
In Section V, we further obtain approximate solutions of the RGE
and provide some numerical examples.  
We then summarize the work in Section VI.

\section{Rephasing invariant parametrization and its properties}  

The rephasing invariant combinations of elements $V_{ij}$ for the neutrino mixing matrix
$V$ (with $\mbox{det}V=+1$) can be constructed by 
the product \cite{Kuo:05,CKL,CK10,CK11,Chiu:2012uc}
\begin{equation}\label{gamma}
\Gamma_{ijk}=V_{1i}V_{2j}V_{3k}=R_{ijk}-iJ,
\end{equation}
where the common imaginary part is identified with the Jarlskog invariant \cite{Jar:85},
and the real parts are defined as
\begin{equation}
(R_{123},R_{231},R_{312};R_{132},R_{213},R_{321})=(x_{1},x_{2},x_{3};y_{1},y_{2},y_{3}).
\end{equation}
The $(x_{i},y_{j})$ variables are constrained by two conditions:
\begin{equation}\label{con1}
det V=(x_{1}+x_{2}+x_{3})-(y_{1}+y_{2}+y_{3})=1,
\end{equation}
\begin{equation}\label{con2}
(x_{1}x_{2}+x_{2}x_{3}+x_{3}x_{1})-(y_{1}y_{2}+y_{2}y_{3}+y_{3}y_{1})=0,
\end{equation}
and they are related to the Jarlskog invariant,
\begin{equation}
J^{2}=x_{1}x_{2}x_{3}-y_{1}y_{2}y_{3}.
\end{equation}
In addition, the $(x_{i},y_{j})$ variables are bounded by $\pm 1$: $-1 \leq (x_{i},y_{j}) \leq +1$,
with $x_{i} \geq y_{j}$ for any pair of $(i,j)$.  

It is convenient to write $|V_{ij}|^{2}$ in a matrix form with elements $x_{i}-y_{j}$:
\begin{equation}
W=[|V_{\alpha i}|^{2}]=
\left(\begin{array}{ccc}
   x_{1}-y_{1} & x_{2}-y_{2} & x_{3}-y_{3} \\
   x_{3}-y_{2} & x_{1}-y_{3} & x_{2}-y_{1} \\
    x_{2}-y_{3} & x_{3}-y_{1} &x_{1}-y_{2} \\
    \end{array}
    \right)
    \end{equation}
The matrix of the cofactors of $W$, denoted as $w$ with $w^{T}W= (\mbox{det}W)I$, is given by
\begin{equation}
w=[|V_{\alpha i}|^{2}]=
\left(\begin{array}{ccc}
   x_{1}+y_{1} & x_{2}+y_{2} & x_{3}+y_{3} \\
   x_{3}+y_{2} & x_{1}+y_{3} & x_{2}+y_{1} \\
    x_{2}+y_{3} & x_{3}+y_{1} &x_{1}+y_{2} \\
    \end{array}
    \right)
    \end{equation}
The elements of $w$ are also bounded, $-1 \leq w_{\alpha i} \leq +1$, and
\begin{equation}
\sum_{i}w_{\alpha i}=\sum_{\alpha}w_{\alpha i}=\mbox{det} W,
\end{equation}
\begin{equation}
\mbox{det} W=\sum x_{i}^{2}-\sum y_{j}^{2}=\sum x_{i}+\sum y_{j}.
\end{equation}   
Note that the constraint equations, Eq.~(\ref{con1}) and Eq.~(\ref{con2}), have been used here.
The relations between this $(x_{i},y_{j})$ parametrization and the standard ones,
$\theta_{12}$, $\theta_{23}$, $\theta_{13}$,
and the Dirac CP phase $\delta$ are shown in Appendix A.

One may further obtain useful expressions of the rephasing invariant combination formed by
products of four mixing elements \cite{Jar:85}, 
\begin{equation}\label{eq:piij}
\pi_{ij}^{\alpha \beta}=V_{\alpha i}V_{\beta j}V_{\alpha j}^{*}V_{\beta i}^{*},
\end{equation}
which can be reduced to 
\begin{eqnarray}
\pi_{ij}^{\alpha \beta} & = & |V_{\alpha i}|^{2}|V_{\beta j}|^{2}-
 \sum_{\gamma k}\epsilon _{\alpha \beta \gamma}\epsilon_{ijk}V_{\alpha i}V_{\beta j}V_{\gamma k} \nonumber \\
   & = & |V_{\alpha j}|^{2}|V_{\beta i}|^{2}+
 \sum_{\gamma k}\epsilon _{\alpha \beta \gamma}\epsilon_{ijk}V_{\alpha j}^{*}V_{\beta i}^{*}V_{\gamma k}^{*},
 \end{eqnarray}
where the second term in either expression is one of the $\Gamma$'s ($\Gamma^{*}$'s) defined in Eq.~(\ref{gamma}).   

In addition, the combination of five elements can be written in the following form:
\begin{equation}
\Xi_{\alpha i}=V_{\alpha j}V_{\alpha k}V_{\alpha i}^{*}V_{\beta i}V_{\gamma i}=(y_{m}y_{n}-x_{b}x_{c})+iJ(1-|V_{\alpha i}|^{2}).
\end{equation}
Here if $|V_{\alpha i}|^{2}=x_{a}-y_{l}$, then $b\neq c\neq a$, $m \neq n \neq l$.
This means that if one takes the $\alpha$th row and the $i$th column, complex conjugates
the vertex ($V_{\alpha i}^{*}$), then the product is rephasing invariant and has a well-defined imaginary part.
Certain intriguing properties of $\Xi$ are shown also in Appendix A.


\section{RGEs for the neutrino parameters} 

The RGE for the Hermitian matrix $M\equiv Y_{\nu}^{\dag}Y_{\nu}$, where $Y_{\nu}$ is the neutrino
Yukawa coupling matrix, is given by 
(see, e.g., \cite{Lindner:2005as,Antusch:2001ck,Chankowski:1993tx,Babu:1993qv,
Casas:1999tg,Antusch:2003kp,Haba:1999fk})
\begin{equation}\label{M}
16\pi^{2}\frac{dM}{dt}=\alpha M+P^{\dag}M+MP
\end{equation} 
at the one-loop level. 
Here, $\alpha$ is real and model-independent,  
$P=CY_{l}^{\dag}Y_{l}+C'Y_{\nu}^{\dag}Y_{\nu}$, with model-dependent
coefficients $C$ and $C'$, and $Y_{l}$ is the charged lepton Yukawa matrix.
Following \cite{Lindner:2005as,Antusch:2001ck,Chankowski:1993tx,Babu:1993qv,
Casas:1999tg,Antusch:2003kp,Haba:1999fk}, we will ignore the term
$C'Y_{\nu}^{\dag}Y_{\nu}$ so that
\begin{equation}
P=CY_{l}^{\dag}Y_{l}.
\end{equation}
Equation (~\ref{M}) is very simple in form.  However, since $M$ contains a number
of unphysical parameters, it is necessary to extract its physical parts.
To this end we may choose the basis where $Y_{l}$ is diagonal, for all energy scales.
One is left then to diagonalize $M$ by the mixing matrix $V$:
\begin{equation}
M=V[diag(h^{2}_{1},h^{2}_{2},h^{2}_{3})]V^{\dag},
\end{equation}
where $h^{2}_{i}$ are the eigenvalues of $M$.
Further, one needs to separate the physical, rephasing invariant, parts of $V$.
To do this we will follows the procedure of 
\cite{Lindner:2005as,Antusch:2001ck,Chankowski:1993tx,Babu:1993qv,
Casas:1999tg,Antusch:2003kp}
but regroup the equations for rephasing invariant variables at the end.

The evolution of the mixing matrix $V$ satisfies the relation, 
\begin{equation}\label{eq:T}
dV/dt=VT,
\end{equation}
here the matrix $T$ is anti-Hermitian.
One may define $\mathcal{D}=16\pi^{2}\frac{d}{dt}$ with $t=\ln(\mu/M_{W})$,
and compares the diagonal elements of 
$\mathcal{D}(V[diag(h^{2}_{1},h^{2}_{2},h^{2}_{3})]V^{\dag})$ with 
that of $\mathcal{D}M$ to obtain
\begin{equation}
\mathcal{D}h^{2}_{i}=h_{i}^{2}[\alpha 
+2 C(|V_{1i}|^{2}f_{1}^{2}+|V_{2i}|^{2}f_{2}^{2}+|V_{3i}|^{2})f_{3}^{2}],
\end{equation}
where $f^{2}_{i}$ are the eigenvalues of the matrix $Y_{l}^{\dag}Y_{l}$,
and the $C'$ terms have been ignored.
From the off-diagonal elements, we obtain the expression of $T_{ij}$:
\begin{equation}\label{eq:HP}
T_{ij}=-H_{ij}P'_{ij}/(16\pi^{2}),
\end{equation}
with $P'= V^{\dag}PV$ and
\begin{equation}\label{eq:H}
H_{ij}=\frac{h_{i}^{2}+h_{j}^{2}}{h_{i}^{2}-h_{j}^{2}}.
\end{equation}

\begin{table}\label{tS}
 \begin{center}
 \begin{tabular}{ccc}
   \\     \hline \hline  
\vspace{0.15in}   
  $S_{11}=\left(\begin{array}{ccc}
   0 & 0 & 0 \\
    0 & \Lambda_{22} & -\Lambda_{23} \\
    0 &-\Lambda_{32} &\Lambda_{33} \\
    \end{array}
    \right)$, 
 &  $S_{12}=\left(\begin{array}{ccc}
   0 & 0 & 0 \\
    -\Lambda_{21} & 0 & \Lambda_{23} \\
    \Lambda_{31} & 0 &-\Lambda_{33} \\
    \end{array}
    \right)$, 
 
    &  $S_{13}=\left(\begin{array}{ccc}
   0 & 0 & 0 \\
   \Lambda_{21} & -\Lambda_{22} & 0 \\
    -\Lambda_{31} &\Lambda_{32} & 0 \\
    \end{array}
    \right)$ 
 
 \\ \vspace{.15in} 
   $S_{21}=\left(\begin{array}{ccc}
   0 & -\Lambda_{12} & \Lambda_{13} \\
   0 & 0 & 0 \\
    0 &\Lambda_{32} &-\Lambda_{33} \\
    \end{array}
    \right)$,    
  &  $S_{22}=\left(\begin{array}{ccc}
   \Lambda_{11} & 0 & -\Lambda_{13} \\
    0 & 0 & 0 \\
    -\Lambda_{31} & 0 &\Lambda_{33} \\
    \end{array}
    \right)$,  
   
  &  $S_{23}=\left(\begin{array}{ccc}
   -\Lambda_{11} & \Lambda_{12} & 0 \\
    0 & 0 & 0 \\
    \Lambda_{31} &-\Lambda_{32} &0 \\
    \end{array}
    \right)$ 
 
 \\ \vspace{.15in} 
   $S_{31}=\left(\begin{array}{ccc}
   0 & \Lambda_{12} & -\Lambda_{13} \\
    0 & -\Lambda_{22} & \Lambda_{23} \\
    0 & 0 & 0 \\
    \end{array}
    \right)$,   
  &  $S_{32}=\left(\begin{array}{ccc}
   -\Lambda_{11} & 0 & \Lambda_{13} \\
    \Lambda_{21} & 0 & -\Lambda_{23} \\
    0 & 0 & 0 \\
    \end{array}
    \right)$, 
   
  &  $S_{33}=\left(\begin{array}{ccc}
   \Lambda_{11} & -\Lambda_{12} & 0 \\
   -\Lambda_{21} & \Lambda_{22} & 0 \\
    0 & 0 & 0 \\
    \end{array}
    \right)$    
   \\  \hline \hline
   
    \end{tabular}
 \caption{The explicit expressions of the matrix $[S_{ij}]$. 
Here $\Lambda _{\gamma k}$ 
is defined in Eq.~(\ref{lambda}).} 
   \end{center}
 \end{table}


To derive the RGE for neutrino mixing parameters, we may start with 
\begin{equation}\label{eq:gamma}
\mathcal{D}\Gamma_{ijk}=\mathcal{D}(V_{1i}V_{2j}V_{3k})
=(\mathcal{D}V_{1i})V_{2j}V_{3k}+V_{1i}(\mathcal{D}V_{2j})V_{3k}+V_{1i}V_{2j}(\mathcal{D}V_{3k}).
\end{equation}
By using Eq.~(\ref{eq:T}), Eq.~(\ref{eq:HP}), and Eq.~(\ref{eq:H}),
one reaches the following form,
\begin{equation}
\mathcal{D}\Gamma_{ijk}=-[(\sum_{l\neq i}V_{1l}H_{li}P'_{li})V_{2j}V_{3k}+
V_{1i}(\sum_{l\neq j}V_{2l}H_{lj}P'_{lj})V_{3k}+V_{1i}V_{2j}(\sum_{l\neq k}V_{3l}H_{lk}P'_{lk})].
\end{equation}
The real part of $\mathcal{D}\Gamma_{ijk}$ gives rise to $\mathcal{D}x_{i}$ and 
$\mathcal{D}y_{i}$ in the following matrix forms:
\begin{equation}\label{eq:Dxi}
\mathcal{D}x_{i}=-C[\Delta f^{2}][A_{i}][H]^{T},
\end{equation}
\begin{equation}\label{eq:Dyi}
\mathcal{D}y_{i}=-C[\Delta f^{2}][A_{i}'][H]^{T},
\end{equation}
where we define 
\begin{equation}
[\Delta f^{2}]=[\Delta f^{2}_{23},\Delta f^{2}_{31},\Delta f^{2}_{12}], 
\end{equation}
\begin{equation}
[H]= [H_{23},H_{31},H_{12}].
\end{equation}
Here $\Delta f^{2}_{ij}=f^{2}_{i}-f^{2}_{j}$.
We have taken over the results of Ref.\cite{Chiu:2008ye}, 
which are also valid for the Dirac neutrino problem.
Thus, the matrices  
matrices $[A_{i}]$ and $[A'_{i}]$ are given in Table II of Ref.~\cite{Chiu:2008ye},
and are reproduced here as Table III in Appendix B.
Note that, to compensate for the usual convention of the neutrino mixing matrix,
whereby it corresponds to the conjugate of that for the quarks, we have adapted
the results of Ref.\cite{Chiu:2008ye} by making the correspondences
charged leptons $\leftrightarrow$ u-type quarks and neutrinos $\leftrightarrow$ d-type quarks.
In addition, the imaginary part of $\mathcal{D}\Gamma_{ijk}$ leads to
\begin{equation}\label{J2}
\mathcal{D}\ln J^{2}=-2C[\Delta f^{2}][w][H]^{T}
\end{equation}



 \begin{table}\label{tZ}
 \begin{center}
 \begin{tabular}{ccc}
   \\     \hline \hline  
\vspace{0.15in}   
  $Z_{1}=\left(\begin{array}{ccc}
   \Lambda_{11} & 0 & 0 \\
    0 & \Lambda_{22} & 0 \\
    0 &0 &\Lambda_{33} \\
    \end{array}
    \right)$, 
 &  $Z_{2}=\left(\begin{array}{ccc}
   0 & \Lambda_{12} & 0 \\
    0 & 0 & \Lambda_{23} \\
    \Lambda_{31}& 0 &0 \\
    \end{array}
    \right)$, 
 
    &  $Z_{3}=\left(\begin{array}{ccc}
   0 & 0 & \Lambda_{13} \\
   \Lambda_{21} & 0 & 0 \\
    0 &\Lambda_{32} & 0 \\
    \end{array}
    \right)$ 
 
 \\ \vspace{.15in} 
   $Z'_{1}=\left(\begin{array}{ccc}
   \Lambda_{11} & 0 & 0 \\
   0 & 0 & \Lambda_{23} \\
    0 &\Lambda_{32} &0 \\
    \end{array}
    \right)$,    
  &  $Z'_{2}=\left(\begin{array}{ccc}
   0 & \Lambda_{12} & 0 \\
    \Lambda_{21} & 0 & 0 \\
   0 & 0 &\Lambda_{33} \\
    \end{array}
    \right)$,  
   
  &  $Z'_{3}=\left(\begin{array}{ccc}
   0 & 0 & \Lambda_{13} \\
    0 & \Lambda_{22} & 0 \\
    \Lambda_{31} &0 &0 \\
    \end{array}
    \right)$     
   \\  \hline 
   \end{tabular}
  \begin{tabular}{ccc}
\vspace{0.25in}

  $[Z_{0}]$ & = & $\left(\begin{array}{ccc}
   (1-|V_{11}|^{2})\Lambda_{11} & (1-|V_{12}|^{2})\Lambda_{12} & (1-|V_{13}|^{2})\Lambda_{13} \\
    (1-|V_{21}|^{2})\Lambda_{21} & (1-|V_{22}|^{2})\Lambda_{22} & (1-|V_{23}|^{2})\Lambda_{23} \\
    (1-|V_{31}|^{2})\Lambda_{31} & (1-|V_{32}|^{2})\Lambda_{32} & (1-|V_{33}|^{2})\Lambda_{33} \\
    \end{array}
    \right)$  \\   

      &=& $\left(\begin{array}{ccc}
   x_{2}x_{3}+y_{2}y_{3}
    & x_{1}x_{3}+y_{1}y_{3} & x_{1}x_{2}+y_{1}y_{2} \\
    x_{1}x_{2}+y_{1}y_{3} & x_{2}x_{3}+y_{1}y_{2} & x_{1}x_{3}+y_{2}y_{3} \\
    x_{1}x_{3}+y_{1}y_{2} & x_{1}x_{2}+y_{2}y_{3} & x_{2}x_{3}+y_{1}y_{3} \\
    \end{array}
    \right)$

   \\  \hline \hline
   
    \end{tabular}
 \caption{The explicit expressions of the matrices $[Z_{i}]$, $[Z'_{i}]$,
 and two equivalent forms of $[Z_{0}]$.} 
   \end{center}
 \end{table}  


It should be noted that, since $\sum \Delta f^{2}_{ij}=0$, the evolution
equations are invariant when a constant is added to any column of
$[A_{i}]$ or $[A'_{i}]$.  For instance,
\begin{equation}\label{eq:Ai}
A_{i} \rightarrow A_{i}+\left(\begin{array}{ccc}
   \delta_{1} & \delta_{2} & \delta_{3} \\
   \delta_{1} &\delta_{2} & \delta_{3} \\
    \delta_{1} & \delta_{2} & \delta_{3} \\
    \end{array}
    \right)
\end{equation}
leaves Eq.~(\ref{eq:Dxi}) invariant.

It turns out that we may recast Eqs.~(\ref{eq:Dxi}) and ~(\ref{eq:Dyi})
into a more symmetrical and suggestive form.  To do that we start with
Eq.~(\ref{eq:piij}) and separate its real and imaginary parts, for $\alpha \neq \beta \neq \gamma$,
$i\neq j \neq k$,
\begin{equation}\label{piab}
\pi^{\alpha \beta}_{ij}\equiv \pi_{\gamma k}=\Lambda_{\gamma k}+iJ.
\end{equation}
Since $Re(\pi^{\alpha \beta}_{ij})$ takes the following forms,
\begin{equation}
Re(\pi^{\alpha \beta}_{ij})=|V_{\alpha i}|^{2}|V_{\beta j}|^{2}-x_{a}=
|V_{\beta i}|^{2}|V_{\alpha j}|^{2}+y_{b},
\end{equation}
we have
\begin{equation}\label{lambda}
\Lambda_{\gamma k}=
\frac{1}{2}(|V_{\alpha i}|^{2}|V_{\beta j}|^{2}+|V_{\alpha j}|^{2}|V_{\beta i}|^{2}-|V_{\gamma k}|^{2}).
\end{equation}
In terms of the $(x,y)$ variables, we find
\begin{equation}\label{lambdaxy}
\Lambda_{\gamma k}=x_{a}y_{j}+x_{b}x_{c}-y_{j}(y_{k}+y_{l}),
\end{equation}
where $(x_{a},y_{j})$ comes from $|V_{\gamma k}|^{2}=x_{a}-y_{j}$, 
and $a\neq b\neq c$, $j\neq k \neq l$.
Another useful identity is
\begin{equation}\label{WLambda}
W_{\gamma k}\Lambda_{\gamma k}=J^{2}+x_{a}y_{j} \hspace{0.3in} \mbox{(no sum)}.
\end{equation}
Using  Eqs.~(\ref{eq:Ai}) and ~(\ref{lambdaxy}), we may simplify the matrices
$[A_{i}]-[A'_{i}]$.  And, with Eqs.~(\ref{eq:Dxi}) and ~(\ref{eq:Dyi}), we find
\begin{equation}\label{vij}
\mathcal{D}W_{ij}=
-2C[\Delta f^{2}][S_{ij}][H]^{T},
\end{equation}
By expressing $(x,y)$ in terms of $W_{ij}$, we further obtain
\begin{equation}\label{x}
\mathcal{D}x_{i}=-C[\Delta f^{2}](2[Z_{i}]-[Z_{0}])[H]^{T},
\end{equation}
\begin{equation}\label{y}
\mathcal{D}y_{i}=-C[\Delta f^{2}](2[Z'_{i}]-[Z_{0}])[H]^{T}.
\end{equation}
Here, $[S_{ij}]$ are given in Table I, while $[Z_{i}]$, $[Z'_{i}]$, and $[Z_{0}]$
are presented in Table II.  
Note that there are two equivalent forms of $[Z_{0}]$,
by using Eqs.~(\ref{eq:Ai}) and ~(\ref{WLambda}).

To exhibit the structure of these equations, let's write down explicitly the evolution equation of $x_{1}$, e.g.,
\begin{equation}
\mathcal{D}x_{1}=-C[\Delta f^{2}][2\left(\begin{array}{ccc}
   \Lambda_{11} & 0 & 0 \\
    0 & \Lambda_{22} & 0 \\
    0 & 0 & \Lambda_{33} \\
    \end{array}
    \right)-[Z_{0}]] [H]^{T}.
    \end{equation}
Also, for $|V_{11}|^{2}$,
\begin{equation}
\mathcal{D}|V_{11}|^{2}=-2C[\Delta f^{2}]\left(\begin{array}{ccc}
   0 & 0 & 0 \\
   0 & \Lambda_{22} & -\Lambda_{23} \\
    0 & -\Lambda_{32} & \Lambda_{33} \\
    \end{array}
    \right)[H]^{T}.
    \end{equation}
The simple forms of these evolution equations mirror the permutation patterns contained
in the definitions of $(x_{i},y_{j})$, such as $x_{1}=Re(\Gamma_{123})=Re(V_{11}V_{22}V_{33})$.
It is also noticeable that $\Lambda_{\alpha i}$, 
which are the real parts of the Jarlskog variables, $\pi^{\beta \gamma}_{jk}$ (Eq.~(\ref{piab})),
and are directly measurable in neutrino
oscillation experiments, play such a prominent role in these evolution equations.
An instructive comparison can be made to the two flavor problem.
Here, the familiar $2 \times 2$ (real) mixing matrix can be parametrized as
$x=C_{\theta}^{2}$, $y=-S_{\theta}^{2}$ and 
$\Lambda=(V_{11}V_{12}V_{21}V_{22})=-C_{\theta}^{2}S_{\theta}^{2}$.
Let us adapt the results (for Majorana neutrinos) of Ref.\cite{Kuo:2001ha}.
With real mass matrices (and $m \rightarrow m^{2}$),
Eq.(29) in Ref.\cite{Kuo:2001ha} becomes
\begin{eqnarray}
\frac{dC_{\theta}^{2}}{dt}=\frac{dx}{dt}&=&
-\chi S^{2}_{2\theta}\frac{m_{2}^{2}+m_{1}^{2}}{m_{2}^{2}-m_{1}^{2}} \nonumber \\
&=&(\Delta f_{21}^{2})\Lambda H_{21}.
\end{eqnarray}
Thus, there is a clear lineage between Eqs.(~\ref{x})-(~\ref{y}) and the two flavor RGE.
It is interesting, and somewhat surprising, that, allowing for the
effects of permutation, the two flavor RGE act simply like building blocks for the
three flavor RGE.

With the above alternative expressions, we may verify that
\begin{equation}
\mathcal{D}(x_{1}+x_{2}+x_{3})=\mathcal(y_{1}+y_{2}+y_{3}),
\end{equation}
\begin{equation}
\mathcal{D}(x_{1}x_{2}+x_{2}x_{3}+x_{3}x_{1})-\mathcal{D}(y_{1}y_{2}+y_{2}y_{3}+y_{3}y_{1})=0,
\end{equation}
and
\begin{equation}
\mathcal{D}J^{2}=\mathcal{D}(x_{1}x_{2}x_{3}-y_{1}y_{2}y_{3}).
\end{equation}

For completeness, we also present the evolution equations for 
the elements of the cofactor matrix $[w]$:
\begin{equation}\label{eq:Dwij}
\mathcal{D}w_{ij}=-2C[\Delta f^{2}][G_{ij}][H]^{T},
\end{equation}
where we have used 
$w_{\gamma k}=|V_{\alpha i}|^{2}|V_{\beta j}|^{2}-|V_{\alpha j}|^{2}|V_{\beta i}|^{2}$
to obtain the matrix $[G_{ij}]$.  They are listed explicitly in
Table IV of Appendix B.


\section{The RGE invariants}

RGE describe the evolution of a multitude of variables as functions of a single parameter, $t$.
As such one might find combinations of physical variables which become independent of $t$.
These RGE invariant prescribe correlations among physical variables and can serve as
powerful constraints on possible theories at high energy.  They have also drawn certain
attention in recent literature \cite{Kuo:2001ha,Chang:2002yr,
Demir:2004aq,Harrison:2010mt,Feldmann:2015nia}).   
To search for neutrino RGE invariants using our parametrization, 
and to further pave the way for the analytic, approximate solutions for $W_{ij}$,
we first define the neutrino mass ratio
$r_{ij}=m_{i}/m_{j}$, where $m_{i}=h_{i}v/\sqrt{2}$, $v\simeq 246$ GeV.
Note that
\begin{equation}
\sinh^{2}\ln r_{ij}=\frac{1}{4}\frac{(m_{i}^{2}-m_{j}^{2})^{2}}{m_{i}^{2}m_{j}^{2}},
\end{equation}
\begin{equation}
\mathcal{D}(\sinh^{2}\ln r_{ij})=
\frac{1}{4}\frac{m_{i}^{4}-m_{j}^{4}}{m_{i}^{2}m_{j}^{2}}(\mathcal{D}\ln m_{i}^{2}-\mathcal{D}\ln m_{j}^{2}),
\end{equation}
and
\begin{equation}
\mathcal{D}[\ln(\sinh^{2}\ln r_{ij})]=
\frac{m_{i}^{2}+m_{j}^{2}}{m_{i}^{2}-m_{j}^{2}}(\mathcal{D}\ln m_{i}^{2}-\mathcal{D}\ln m_{j}^{2}).
\end{equation}
This leads to
\begin{equation}
\Sigma_{ij}\mathcal{D}[\ln(\sinh^{2}\ln r_{ij})]=
2C[\Delta f^{2}][w][H]^{T}.
\end{equation}
Combining this result with the expression of $\mathcal{D}\ln J^{2}$, we find that
\begin{equation}\label{inv-1}
\mathcal{D}[\ln [ J^{2}(\sinh^{2}\ln r_{12})(\sinh^{2}\ln r_{23})(\sinh^{2}\ln r_{31})]]=0,
\end{equation}
i.e., $J^{2}(\Pi_{ij}\sinh^{2}\ln r_{ij})$
is a RGE invariant.

Furthermore, one notes that any quantity that is identical to $J^{2}$ also gives 
rise to an invariant when
multiplied by $\Pi_{ij}\sinh^{2}\ln r_{ij}$ in Eq.~(\ref{inv-1}).
With the expression of $\Lambda_{\gamma k}$ in Eq.~(\ref{lambda}),
it is straightforward to write down nine different forms of $J^{2}=\pi^{2}_{\gamma k}-\Lambda^{2}_{\gamma k}$,
which correspond to nine different combinations of $(\gamma,k)$,
\begin{eqnarray}
J^{2}& = & \pi^{2}_{\gamma k}-\Lambda^{2}_{\gamma k} \nonumber \\
   & = &  |V_{\alpha i}|^{2}|V_{\beta j}|^{2}|V_{\alpha j}|^{2}|V_{\beta i}|^{2}-\Lambda^{2}_{\gamma k}.
\end{eqnarray}
This leads to nine RGE invariants which consist of $|V_{ij}|^{2}$ and the mass ratios $\ln r_{ij}^{2}$:
\begin{equation}\label{inv}
(|V_{\alpha i}|^{2}|V_{\beta j}|^{2}|V_{\alpha j}|^{2}|V_{\beta i}|^{2}-\Lambda^{2}_{\gamma k})
(\sinh\ln r_{21}^{2})(\sinh\ln r_{32}^{2})
(\sinh\ln r_{13}^{2})=\mbox{constant}
\end{equation}

We may also study approximate solutions of Eqs.~(\ref{J2}),~(\ref{vij}).
Consider the familiar case of hierarchy, $f^{2}_{3} \gg (f^{2}_{2},f^{2}_{1})$, $[\Delta f^{2}]\cong f^{2}_{3}[-1,+1,0]$.
Then
\begin{eqnarray}\label{lnJ}
\frac{1}{2C}\mathcal{D}\ln J^{2} & \cong & [(w_{11}-w_{21}),(w_{12}-w_{22}),(w_{13}-w_{32})][H]^{T}  \nonumber \\ 
            &=& [(|V_{33}|^{2}-|V_{32}|^{2}),(|V_{31}|^{2}-|V_{33}|^{2}),(|V_{32}|^{2}-|V_{31}|^{2})][H]^{T}.
\end{eqnarray}
Under the same approximation, we find, e.g.,
\begin{eqnarray}\label{appv}
\frac{1}{2C}\mathcal{D}\ln |V_{31}|^{2} & \cong & \frac{f^{2}_{3}}{|V_{31}|^{2}}[1,-1.0]
\left(\begin{array}{ccc}
   0 & \Lambda_{12} & -\Lambda_{13} \\
    0 & -\Lambda_{22} & \Lambda_{23} \\
    0 & 0 & 0 \\
    \end{array}
    \right)[H]^{T} \nonumber \\
           & = & f^{2}_{3}[0,-|V_{33}|^{2},|V_{32}|^{2}][H]^{T},
 \end{eqnarray}
 where we used the relations $\Lambda _{12}+\Lambda_{22}=-|V_{31}|^{2}|V_{33}|^{2}$
 and $\Lambda_{13}+\Lambda_{23}=-|V_{31}|^{2}|V_{32}|^{2}$.
 Together with similar results for $\mathcal{D}\ln |V_{22}|^{2}$ and $\mathcal{D} \ln|V_{33}|^{2}$,
 we find the following approximate RGE invariant:
 \begin{equation}
 J^{2}/(|V_{31}|^{2}|V_{32}|^{2}V_{33}|^{2})\cong \mbox{invariant}, \\ f^{2}_{3}\gg (f^{2}_{2},f^{2}_{1}).
 \end{equation}
 The approximate solutions, Eqs.~(\ref{lnJ}),~(\ref{appv}), are actually quite accurate as long as
 the hierarchy used is not upended by renormalization.


Furthermore, if the neutrino masses satisfy the hierarchical condition: $H_{ij} \gg H_{jk},H_{ki}$,
then 
\begin{equation}
\frac{d \ln J^{2}}{dt} + \frac{d \ln(\sinh^{2}\ln r_{ij})}{dt} \simeq 0
\end{equation}
and the approximate invariant follows:
\begin{equation}
 J^{2}(\sinh^{2}\ln r_{ij})=\mbox{constant}.
 \end{equation}
Since $H_{12} \gg H_{23},H_{31}$ in general, we may write
\begin{equation}
 J^{2}(\sinh^{2}\ln r_{12})=\mbox{constant}.
 \end{equation}



\section{Approximate solutions}

Even though complete analytical solutions to the coupled RGE are unavailable,
certain approximations to the RGE may lead us to
solutions that are surprisingly simple.
We first consider a hierarchical scenario for the charged leptons, 
$f^{2}_{3} \gg f^{2}_{2}$, $f^{2}_{1}$, which yields
\begin{equation}\label{eq:f}
[\Delta f^{2}_{23},\Delta f^{2}_{31},\Delta f^{2}_{12}] \approx f^{2}_{3}[-1,1,0].
\end{equation}
In addition, we have 
\begin{equation}\label{eq:H}
[H_{23},H_{31},H_{12}] \approx [-1,1,-1]
\end{equation}
if the neutrino masses are hierarchical: $h^{2}_{3} \gg h^{2}_{2} \gg h^{2}_{1}$.
Substituting Eqs.~(\ref{eq:f}) and ~(\ref{eq:H}) in Eq.~(\ref{vij})
results in the following simple forms:
\begin{equation}\label{DW11}
C'\mathcal{D}W_{11}=-(\Lambda_{22}+\Lambda_{23})=W_{11}W_{31},
\end{equation}
\begin{equation}
C'\mathcal{D}W_{12}=-\Lambda_{21}+\Lambda_{23},
\end{equation}
\begin{equation}
C'\mathcal{D}W_{13}=\Lambda_{21}+\Lambda_{22}=-W_{13}W_{33},
\end{equation}
\begin{equation}
C'\mathcal{D}W_{21}=-(\Lambda_{12}+\Lambda_{13})=W_{21}W_{31},
\end{equation}
\begin{equation}
C'\mathcal{D}W_{22}=-\Lambda_{11}+\Lambda_{13},
\end{equation}
\begin{equation}
C'\mathcal{D}W_{23}=\Lambda_{11}+\Lambda_{12}=-W_{23}W_{33},
\end{equation}
\begin{equation}\label{DW31}
C'\mathcal{D}W_{31}=\Lambda_{12}+\Lambda_{22}+\Lambda_{13}+\Lambda_{23}=-W_{31}(1-W_{31}),
\end{equation}
\begin{equation}
C'\mathcal{D}W_{32}=(\Lambda_{11}+\Lambda_{21})-(\Lambda_{13}+\Lambda_{23})=W_{32}(W_{31}-W_{33}),
\end{equation}
\begin{equation}\label{DW33}
C'\mathcal{D}W_{33}=-(\Lambda_{11}+\Lambda_{21})-(\Lambda_{12}+\Lambda_{22})=W_{33}(1-W_{33}),
\end{equation}
where $C'=1/(2Cf^{2}_{3})$.
One observes that the six quantities, 
$\mathcal{D}\ln W_{11}$, $\mathcal{D}\ln W_{13}$, $\mathcal{D}\ln W_{21}$, $\mathcal{D}\ln W_{23}$,
$\mathcal{D}\ln W_{31}$, and $\mathcal{D}\ln W_{33}$ depend only on $W_{31}$ and $W_{33}$.
It is also seen that the following two relations hold:
\begin{equation}
\mathcal{D}\ln W_{13}=\mathcal{D}\ln W_{23},
\end{equation}
\begin{equation}
\mathcal{D}\ln W_{11}=\mathcal{D}\ln W_{21}.
\end{equation}

\begin{figure}[ttt]
\caption{Evolution of $|V_{11}|^{2}$ (left column) and $|V_{13}|^{2}$ (right column)
from high to low energies. 
The approximate solutions (solid) are evaluated using constant $f^{2}_{3}$: 
$f^{2}_{3}=10^{-4}$, $f^{2}_{3}=10^{-2}$, and $f^{2}_{3}=1$.
The full solutions (dashed) are also plotted with the initial values,
$f^{2}_{3}=10^{-4}$, $f^{2}_{3}=10^{-2}$, and $f^{2}_{3}=1$
at $t_{0}=30$.} 
\centerline{\epsfig{file=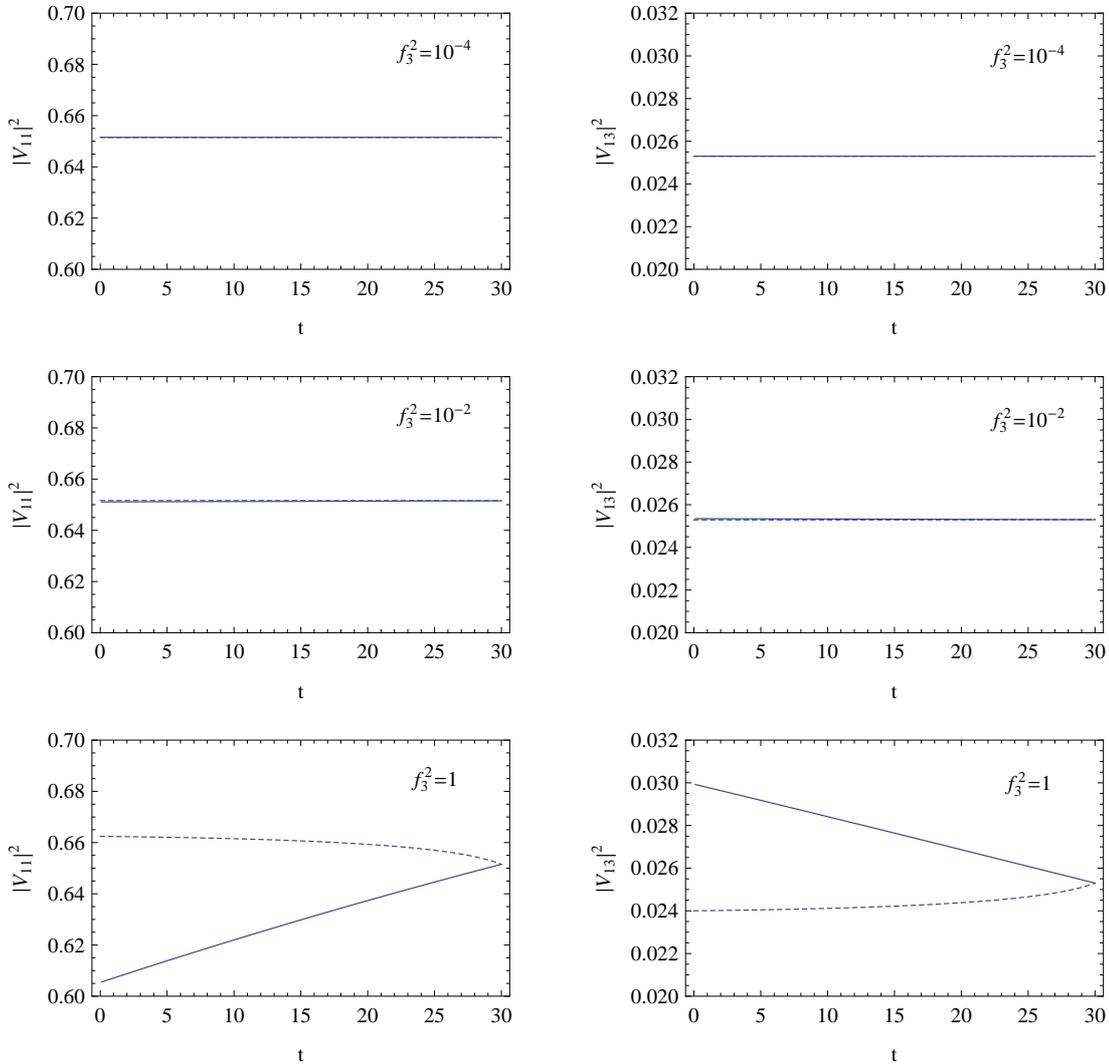,width= 15cm}}
\end{figure}

One notes that Eq.~(\ref{DW31}) has an approximate, analytic solution for a
range of $t$-values inside which 
$f^{2}_{3}\approx$constant is a good approximation:
\begin{equation}\label{eq:W31}
W_{31}\cong \frac{1}{(a_{31}^{-1}-1)e^{(t-t_{0})/C''}+1},
\end{equation}
where $C''=16\pi^{2}C'$, and $a_{ij}$ is the initial value of $W_{ij}$ at $t=t_{0}$.
Similarly, we can solve for $W_{33}$ from Eq.~(\ref{DW33}):
 \begin{equation}\label{eq:W33}
W_{33}\cong \frac{1}{(a_{33}^{-1}-1)e^{-(t-t_{0})/C''}+1}.
\end{equation}
With the explicit expressions for $W_{31}$ in Eq.~(\ref{eq:W31}),
we may rewrite Eq.~(\ref{DW11}) as:
\begin{equation}
C'\ln W_{11}=\int \frac{dt}{(a_{31}^{-1}-1)e^{-(t-t_{0})/C''}+1},
\end{equation}
which can be solved for $W_{11}$:
\begin{equation}
W_{11}\cong \frac{a_{11}}{(1-a_{31})+a_{31}e^{-(t-t_{0})/C''}}.
\end{equation}
Similarly, $W_{13}$ can be solved by using the explicit solution of $W_{33}$ in Eq.~(\ref{eq:W33}),
\begin{equation}
W_{13}\cong \frac{a_{13}}{(1-a_{33})+a_{33}e^{(t-t_{0})/C''}}.
\end{equation}
The rest of $W_{ij}$ can be obtained directly by using $\sum_{i}W_{ij}=1$
and $\sum_{j}W_{ij}=1$.

A confidence builder for these approximate solutions is the numerical solutions of RGE with initial
conditions incorporating the approximations in Eqs.~(\ref{eq:f}) and ~(\ref{eq:H}).
Note that
to assess the general nature of the RGE, 
it seems more appropriate to start from a point with fast evolution, 
so that most changes may be accomplished in its neighborhood,
with minor corrections afterwards.
The low energy physics values are close to a fixed point
of the RGE, and it is crucial to study how they are approached from the high energy values.   
However, without detailed knowledge of the initial values at high energy in the scope of 
theoretical framework, it is difficult to assign a suitable region 
of initial parameter space that yields all the measured values
as the RGE evolve down to the low energy.

At high energy, either hierarchical or 
near degenerate masses and mixing parameters are both motivated
by various models \cite{Mohapatra:2003tw}. For the purpose of illustration, 
we simply assume $C=-3/2$ in analogy to the quark RGE
and adopt the following initial input parameters at high energy:
\begin{itemize}
\item $f^{2}_{3}=10^{-4}$, $f^{2}_{3}=10^{-2}$, 
and $f^{2}_{3}=1$ are adopted for calculating the approximate solutions,
which will be compared with the full solution using the same values of $f_{3}^{2}$ as
the initial input at $t_{0}=30$.
\item The neutrino masses at $t_{0}=30$ are taken to be hierarchical: 
$h^{2}_{3} \gg h^{2}_{2} \gg h^{2}_{1}$, which leads to Eq.~(\ref{eq:H}).
\item It is found that $(x_{i},y_{j})$ for neutrinos
only evolve slightly from their respective initial values.
We therefore adopt 
$[x_{1},x_{2};y_{1},y_{2}]=[(1/3)-\epsilon,(1/6)-\epsilon;(-1/3)+\epsilon,(-1/6)+\epsilon]$
with $\epsilon=0.01$ as the input at $t_{0}=30$ so as to yield reasonable 
mixing parameters at low energy. 
\end{itemize}

With the above inputs, we present examples of approximate and full numerical solutions
for $|V_{11}|^{2}$ and $|V_{13}|^{2}$ in Fig.1, 
and summarize the results in the following.
\begin{itemize}
\item The approximate solutions agree well with the full solutions for small $f^{2}_{3}$.
However, deviation begins to enlarge for larger $f^{2}_{3}$, e.g., when $f^{2}_{3}\sim 1$.
\item Although the quark mixing elements can evolve quite significantly from
high energy \cite{Chiu:2008ye}, 
the neutrino mixing parameters ($x_{i}$ and $y_{i}$) only evolve slightly
and thus the elements $W_{ij}=|V_{ij}|^{2}$ do not evolve much from the 
initial values.  
This general behavior is in agreement with the expectation of 
recent study (see, e.g., Ref\cite{Ohlsson:2013xva}).  
\item Note that with the chosen input values, some of the parameters do not
evolve to the observed values at low energies. It appears that the RGE evolution may be 
sensitive to the initial parameters
and that fine-tuning the may be required. 
\end{itemize}

Finally, it is worthwhile to mention that the RGE for $W_{ij}=|V_{ij}|^{2}$ may also be simplified
into different forms if $f_{i}^{2}$ remain hierarchical: $[\Delta f^{2}] \approx f^{2}_{3}[-1,1,0]$,
while a pair of $h^{2}_{i}$'s are now nearly degenerate: 
$h^{2}_{3}\geq h^{2}_{2} \approx h^{2}_{1}$.
It leads to 
\begin{equation}
[H]\approx (\frac{h^{2}_{1}+h^{2}_{2}}{h^{2}_{1}-h^{2}_{2}})[0,0,1]=H_{12}[0,0,1].
\end{equation}
In this case, we obtain the following simple forms for $\mathcal{D}W_{ij}$:
\begin{equation} 
\bar{C}\mathcal{D}W_{11}=-\Lambda_{23},
\end{equation}
\begin{equation} 
\bar{C}\mathcal{D}W_{12}=\Lambda_{23},
\end{equation}
\begin{equation} 
\bar{C}\mathcal{D}W_{13}=0,
\end{equation}
\begin{equation} 
\bar{C}\mathcal{D}W_{21}=-\Lambda_{13},
\end{equation}
\begin{equation} 
\bar{C}\mathcal{D}W_{22}=\Lambda_{13},
\end{equation}
\begin{equation} 
\bar{C}\mathcal{D}W_{23}=0,
\end{equation}
\begin{equation} 
\bar{C}\mathcal{D}W_{31}=\Lambda_{13}+\Lambda_{23},
\end{equation}
\begin{equation} 
\bar{C}\mathcal{D}W_{32}=-(\Lambda_{13}+\Lambda_{23}),
\end{equation}
\begin{equation} 
\bar{C}\mathcal{D}W_{33}=0,
\end{equation}
where $\bar{C}=C'/H_{12}$.

The RGE (Eq.~(\ref{x})-~(\ref{y})) are valid for a class of theories that
are similar to the SM.  The examples that we presented also show that these
equations have simple approximate solutions under various sets of input parameters.
In the search of theories beyond the standard model, they should be helpful
to identify those that can yield viable sets of physical parameters at low energies.

\section{conclusion}

One of the most important questions in the study of neutrinos 
is an assessment of the RGE evolution of neutrino masses and mixing.
In contrast to the corresponding problem for gauge couplings, owing to
the complexity of the RGE, the evolution of the neutrino parameters is rather poorly understood.
It is thus not easy to analyze models at high energies which may start with
a variety of mass and/or mixing patterns.
We studied the one-loop RGE of Dirac neutrinos, using a rephasing invariant parametrization
introduced earlier.  Because of the symmetry structure (under flavor permutation)
of this parametrizations, it is found that the resulting equations
(Eqs.~(\ref{J2}),~(\ref{vij}),~(\ref{x}),~(\ref{y})) 
can be arranged into a highly symmetric matrix form,
which can facilitate both theoretical and numerical studies of the solutions.
In particular, they should provide a useful guide in the search
of extensions of the SM.
We obtained a RGE invariant, as well as some approximate solutions.
In addition, some numerical results are presented.

Work is in progress to generalize our results to the case of Majorana neutrinos, as well as the
evolution of the quark mass matrices.  We hope to present these studies in a future publication.


\acknowledgments                 
SHC is supported by the Ministry of Science and Technology of Taiwan, 
Grant No.: MOST 104-2112-M-182-004.

\appendix

\section{}

We first briefly outline the connection between the $(x,y)$ parametrization
and the standard one in the following.
It can be verified that the mixing angles of the standard parametrization:
$s^{2}_{12}\equiv \sin^{2}\theta_{12}$, $s^{2}_{23}\equiv \sin^{2}\theta_{23}$,
and $s^{2}_{13}\equiv \sin^{2}\theta_{13}$ 
are related to the $(x_{i},y_{j})$ parameters,
\begin{equation}
s^{2}_{12}=1/(1+\frac{x_{1}-y_{1}}{x_{2}-y_{2}}),
\end{equation}
\begin{equation}
s^{2}_{23}=1/(1+\frac{x_{1}-y_{2}}{x_{2}-y_{1}}),
\end{equation}
\begin{equation}
s^{2}_{13}=x_{3}-y_{3}.
\end{equation}
On the other hand, one may also show that, with $c_{ij}\equiv \cos\theta_{ij}$
and $K\equiv s_{12}c_{12}s_{13}c^{2}_{13}s_{23}c_{23}$,
\begin{eqnarray}
J&=&K\sin\varphi \nonumber \\
x_{1}&=&c^{2}_{12}c^{2}_{13}c^{2}_{23}-K\cos\delta \nonumber \\
x_{2}&=&s^{2}_{12}c^{2}_{13}s^{2}_{23}-K\cos\delta \nonumber \\
x_{3}&=&s^{2}_{12}s^{2}_{13}c^{2}_{23}+c^{2}_{12}s^{2}_{13}s^{2}_{23}+
      \frac{1+s^{2}_{13}}{1-s^{2}_{13}}K\cos\delta \nonumber \\
y_{1}&=&-c^{2}_{12}c^{2}_{13}s^{2}_{23}-K\cos\delta \nonumber \\
y_{2}&=&-s^{2}_{12}c^{2}_{13}c^{2}_{23}-K\cos\delta \nonumber \\
y_{3}&=&-s^{2}_{12}s^{2}_{13}s^{2}_{23}-c^{2}_{12}s^{2}_{13}c^{2}_{23}+
      \frac{1+s^{2}_{13}}{1-s^{2}_{13}}K\cos\delta,
\end{eqnarray}
where $\delta$ is the Dirac CP phase in the standard parametrization.
Note that $\Xi_{13}=Ke^{i\delta}(1-|V_{13}|^{2})$. Thus, the phase $\delta$
can be identified as the phase of the rephasing invariant quantity $\Xi_{13}$.

In addition, we list some interesting properties of $\Xi$ in the following.
A straightforward calculation shows that 
\begin{equation}\label{A5}
\sum_{\alpha} \Xi_{\alpha i}=\sum_{i} \Xi_{\alpha i}=2iJ.  
\end{equation} 
One may consider the matrix $[\Xi]$ and its cofactor matrix $[\xi]$,
\begin{eqnarray}
[\Xi]=\left(\begin{array}{ccc}
   \Xi_{e1} & \Xi_{e2} & \Xi_{e3} \\
    \Xi_{\mu 1} & \Xi_{\mu 2} & \Xi_{\mu 3} \\
    \Xi_{\tau 1} & \Xi_{\tau 2} & \Xi_{\tau 3} \\
    \end{array},
    \right)
\end{eqnarray}   

   \begin{eqnarray}
[\xi]=\left(\begin{array}{ccc}
   \xi_{e1} & \xi_{e2} & \xi_{e3} \\
    \xi_{\mu 1} & \xi_{\mu 2} & \xi_{\mu 3} \\
    \xi_{\tau 1} & \xi_{\tau 2} & \xi_{\tau 3} \\
    \end{array}
    \right).
\end{eqnarray} 
From Eq.(~\ref{A5}), we see that the matrices $[\Xi]$ and $[\xi]$ have similar properties as the 
pair $W$ and $w$. In particular,
\begin{equation}
det[\Xi]=2iJ\sum_{\alpha}\xi_{\alpha i}=2iJ \sum_{i}\xi_{\alpha i}.
\end{equation}
One may also find the relations between $\Xi_{\alpha i}$ and 
$\pi_{\gamma k}\equiv \pi^{\alpha \beta}_{ij}$, e.g., 
\begin{eqnarray}
\pi_{e1} & = & \Xi_{\tau 2}-(x_{1}-iJ)W_{\tau 2} \nonumber \\
        & = & \Xi_{\mu 3}-(x_{1}-iJ)W_{\mu 3} \nonumber \\
       & = & -\Xi_{\tau 3}^{*}+(y_{1}+iJ)W_{\tau 3} \nonumber \\
       & = & -\Xi_{\mu 2}^{*}+(y_{1}+iJ)W_{\mu 2}.
       \end{eqnarray}


 \section{}      
       
The explicit matric forms of $[A_{i}]$ and $[A'_{i}]$ in 
Eqs.~(\ref{eq:Dxi}) and (\ref{eq:Dyi}) 
are shown in Table III, and that for $G_{ij}$ in Eqs.~(\ref{eq:Dwij})
is shown in Table IV.

 \begin{table}[h]\label{t:A}
 \begin{center}
 \begin{tabular}{cc}
$[A_{i}]$ & $[A'_{i}]$    \\     \hline \hline  
\vspace{0.25in}   
  $[A_{1}]=\left(\begin{array}{ccc}
   2x_{1}y_{1} & x_{1}x_{2}+y_{2}y_{3} & x_{1}x_{3}+y_{2}y_{3} \\
    x_{1}x_{3}+y_{1}y_{2} & 2x_{1}y_{3} & x_{1}x_{2}+y_{1}y_{2} \\
    x_{1}x_{2}+y_{1}y_{3} & x_{1}x_{3}+y_{1}y_{3} & 2x_{1}y_{2} \\
    \end{array}
    \right)$, 
 &  $[A'_{1}]=\left(\begin{array}{ccc}
    2x_{1}y_{1} & x_{2}x_{3}+y_{1}y_{2} & x_{2}x_{3}+y_{1}y_{3} \\
    x_{1}x_{3}+y_{1}y_{2} & x_{1}x_{3}+y_{1}y_{3} & 2x_{2}y_{1} \\
    x_{1}x_{2}+y_{1}y_{3} & 2x_{3}y_{1} & x_{1}x_{2}+y_{1}y_{2} \\
    \end{array}
    \right)$

 \\ \vspace{.25in} 
   $[A_{2}]=\left(\begin{array}{ccc}
    x_{1}x_{2}+y_{1}y_{3}&2x_{2}y_{2} & x_{2}x_{3}+y_{1}y_{3} \\
    x_{2}x_{3}+y_{2}y_{3} & x_{1}x_{2}+y_{2}y_{3} & 2x_{2}y_{1} \\
    2x_{2}y_{3} & x_{2}x_{3}+y_{1}y_{2} & x_{1}x_{2}+y_{1}y_{2} \\
    \end{array}
    \right)$,   
  &  $[A'_{2}]=\left(\begin{array}{ccc}
 x_{1}x_{3}+y_{1}y_{2}  & 2x_{2}y_{2} & x_{1}x_{3}+y_{2}y_{3} \\
   2x_{3}y_{2} & x_{1}x_{2}+y_{2}y_{3} & x_{1}x_{2}+y_{1}y_{2} \\
    x_{2}x_{3}+y_{2}y_{3} & x_{2}x_{3}+y_{1}y_{2} & 2x_{1}y_{2} \\
    \end{array}
    \right)$

 \\ \vspace{.25in}
       
    $[A_{3}]=\left(\begin{array}{ccc}
    x_{1}x_{3}+y_{1}y_{2} & x_{2}x_{3}+y_{1}y_{2} & 2x_{3}y_{3} \\
    2x_{3}y_{2} & x_{1}x_{3}+y_{1}y_{3} & x_{2}x_{3}+y_{1}y_{3} \\
    x_{2}x_{3}+y_{2}y_{3} & 2x_{3}y_{1} & x_{1}x_{3}+y_{2}y_{3} \\
    \end{array}
    \right)$,
 &   $[A'_{3}]=\left(\begin{array}{ccc}
    x_{1}x_{2}+y_{1}y_{3} & x_{1}x_{2}+y_{2}y_{3} & 2x_{3}y_{3} \\
    x_{2}x_{3}+y_{2}y_{3} & 2x_{1}y_{3} & x_{2}x_{3}+y_{1}y_{3} \\
    2x_{2}y_{3} & x_{1}x_{3}+y_{1}y_{3} & x_{1}x_{3}+y_{2}y_{3} \\
    \end{array}
    \right)$  
  
   \\  \hline \hline
   
    \end{tabular}
 \caption{The explicit expressions of the matrices$[A_{i}]$ and $[A'_{i}]$.} 
   \end{center}
 \end{table}


\begin{table}\label{t:G}
 \begin{center}
 \begin{tabular}{c}
   \\     \hline \hline  
\vspace{0.25in}   
  $G_{11}=\left(\begin{array}{ccc}
   -\Lambda_{11}(1+|V_{11}|^{2}) & \Lambda_{12}(1-|V_{12}|^{2}) & \Lambda_{13}(1-|V_{13}|^{2}) \\
    \Lambda_{21}(1-|V_{21}|^{2}) & -\Lambda_{22}|V_{22}|^{2} & -\Lambda_{23}|V_{23}|^{2} \\
    \Lambda_{31}(1-|V_{31}|^{2}) &-\Lambda_{32}|V_{32}|^{2} &-\Lambda_{33}|V_{33}|^{2} \\
    \end{array}
    \right)$
     \\ \vspace{.25in} 
  $G_{12}=\left(\begin{array}{ccc}
   \Lambda_{11}(1-|V_{11}|^{2}) & -\Lambda_{12}(1+|V_{12}|^{2}) & \Lambda_{13}(1-|V_{13}|^{2}) \\
    -\Lambda_{21}|V_{21}|^{2} & \Lambda_{22}(1-|V_{22}|^{2}) & -\Lambda_{23}|V_{23}|^{2} \\
    -\Lambda_{31}|V_{31}|^{2} & \Lambda_{32}(1-|V_{32}|^{2}) &-\Lambda_{33}|V_{33}|^{2} \\
    \end{array}
    \right)$
     \\ \vspace{.25in} 
  $G_{13}=\left(\begin{array}{ccc}
   \Lambda_{11}(1-|V_{11}|^{2}) & \Lambda_{12}(1-|V_{12}|^{2}) & -\Lambda_{13}(1+|V_{13}|^{2}) \\
   -\Lambda_{21}|V_{21}|^{2} & -\Lambda_{22}|V_{22}|^{2} & \Lambda_{23}(1-|V_{23}|^{2}) \\
    -\Lambda_{31}|V_{31}|^{2}&-\Lambda_{32}|V_{32}|^{2} &\Lambda_{33}(1-|V_{33}|^{2}) \\
    \end{array}
    \right)$ 
 \\ \vspace{.25in} 
   $G_{21}=\left(\begin{array}{ccc}
  \Lambda_{11}(1-|V_{11}|^{2}) & -\Lambda_{12}|V_{12}|^{2} & -\Lambda_{13}|V_{13}|^{2} \\
   -\Lambda_{21}(1+|V_{21}|^{2}) & \Lambda_{22}(1-|V_{22}|^{2}) & \Lambda_{23}(1-|V_{23}|^{2}) \\
    \Lambda_{31}(1-|V_{31}|^{2}) &-\Lambda_{32}|V_{32}|^{2} & -\Lambda_{33}|V_{33}|^{2} \\
    \end{array}
    \right)$
    \\ \vspace{.25in}     
    $G_{22}=\left(\begin{array}{ccc}
   -\Lambda_{11}|V_{11}|^{2}  & \Lambda_{12}(1-|V_{12}|^{2}) & -\Lambda_{13}|V_{13}|^{2}  \\
    \Lambda_{21}(1-|V_{21}|^{2}) & -\Lambda_{22}(1+|V_{22}|^{2}) & \Lambda_{23}(1-|V_{23}|^{2}) \\
    -\Lambda_{31}|V_{31}|^{2}  & \Lambda_{32}(1-|V_{32}|^{2}) &-\Lambda_{33}|V_{33}|^{2}  \\
    \end{array}
    \right)$
    \\ \vspace{.25in}   
    $G_{23}=\left(\begin{array}{ccc}
   -\Lambda_{11}|V_{11}|^{2} & -\Lambda_{12}|V_{12}|^{2} &  \Lambda_{13}(1-|V_{13}|^{2}) \\
    \Lambda_{21}(1-|V_{21}|^{2})&  \Lambda_{22}(1-|V_{22}|^{2}) &  -\Lambda_{23}(1+|V_{23}|^{2}) \\
    -\Lambda_{31}|V_{31}|^{2}& -\Lambda_{32}|V_{32}|^{2} & \Lambda_{33}(1-|V_{33}|^{2}) \\
    \end{array}
    \right)$ 
 \\ \vspace{.25in} 
   $G_{31}=\left(\begin{array}{ccc}
   \Lambda_{11}(1-|V_{11}|^{2}) &-\Lambda_{12}|V_{12}|^{2} & -\Lambda_{13}|V_{13}|^{2} \\
    \Lambda_{21}(1-|V_{21}|^{2}) &-\Lambda_{22}|V_{22}|^{2} & -\Lambda_{23}|V_{23}|^{2} \\
    -\Lambda_{31}(1+|V_{31}|^{2}) & \Lambda_{32}(1-|V_{32}|^{2}) & \Lambda_{33}(1-|V_{33}|^{2}) \\
    \end{array}
    \right)$
      \\ \vspace{.25in}  
   $G_{32}=\left(\begin{array}{ccc}
    -\Lambda_{11}|V_{11}|^{2}& \Lambda_{12}(1-|V_{12}|^{2}) & -\Lambda_{13}|V_{13}|^{2}) \\
    -\Lambda_{21}|V_{21}|^{2} & \Lambda_{22}(1-|V_{22}|^{2}) & -\Lambda_{23}|V_{23}|^{2} \\
    -\Lambda_{31}(1-|V_{31}|^{2})&  -\Lambda_{32}(1+|V_{32}|^{2}) &  \Lambda_{33}(1-|V_{33}|^{2}) \\
    \end{array}
    \right)$
     \\ \vspace{.25in} 
     $G_{33}=\left(\begin{array}{ccc}
  -\Lambda_{11}|V_{11}|^{2} & -\Lambda_{12}|V_{12}|^{2} &  \Lambda_{13}|V_{13}|^{2} \\
   -\Lambda_{21}|V_{21}|^{2} & -\Lambda_{22}|V_{22}|^{2} &  \Lambda_{23}(1-|V_{23}|^{2}) \\
    \Lambda_{31}(1+|V_{31}|^{2}) & -\Lambda_{32}(1-|V_{32}|^{2}) & -\Lambda_{33}(1+|V_{33}|^{2}) \\
    \end{array}
    \right)$ 
   \\  \hline \hline
   
    \end{tabular}
 \caption{The explicit expressions of the matrix $[G_{ij}]$. 
Here $\Lambda _{\gamma k}$ 
is defined in Eq.~(\ref{lambda}).} 
   \end{center}
 \end{table}



\end{document}